# menoci: Lightweight Extensible Web Portal enabling FAIR Data Management for Biomedical Research Projects

Suhr M[*,1], Lehmann C[1], Bauer CR[1], Bender T[1], Knopp C[1], Freckmann LK[1], Öst Hansen B[1], Henke C[1], Aschenbrandt G[1], Kühlborn LK[1], Rheinländer S[1], Weber L[1], Marzec B[1], Hellkamp M[2], Wieder P[2], Kusch H[1,3], Sax U[1], Nussbeck SY[1,4]

* corresponding author
[1] University Medical Center Göttingen, Dept. of Medical Informatics, von-Siebold-Str. 3, 37075 Göttingen, Germany
[2] GWDG, Gesellschaft für wissenschaftliche Datenverarbeitung mbH Göttingen, Am Faßberg 11, 37077 Göttingen, Germany
[3] University Medical Center Göttingen, Dept. of Molecular Biology, Humboldtallee 23, 37075 Göttingen, Germany
[4] University Medical Center Göttingen, UMG Biobank, Robert-Koch-Str. 40, 37075 Göttingen, Germany

## Abstract

**Background**: Biomedical research projects deal with data management requirements from multiple sources like funding agencies' guidelines, publisher policies, discipline best practices, and their own users' needs. We describe functional and quality requirements based on many years of experience implementing data management for the CRC 1002 and CRC 1190. A fully equipped data management software should improve documentation of experiments and materials, enable data storage and sharing according to the FAIR Guiding Principles while maximizing usability, information security, as well as software sustainability and reusability.

**Results**: We introduce the modular web portal software *menoci* for data collection, experiment documentation, data publication, sharing, and preservation in biomedical research projects. *Menoci* modules are based on the Drupal content management system which enables lightweight deployment and setup, and creates the possibility to combine research data management with a customisable project home page or collaboration platform.

**Conclusions**: Management of research data and digital research artefacts is transforming from individual researcher or groups best practices towards project- or organisation-wide service infrastructures. To enable and support this structural transformation process, a vital ecosystem of open source software tools is needed. *Menoci* is a contribution to this ecosystem of research data management tools that is specifically designed to support biomedical research projects.

# Background

Emerging data-driven research methods and the push for open and reproducible science amplify the need for strategic approaches to the management of research data across disciplines [1]. Data science and "big data" analytic applications require streamlined data collection and metadata annotation by the initial producers of source data, i.e. researchers and experimentalists in the laboratory considering the biomedical context. Research data management (RDM) is often described as a life cycle spanning task including the planning of a data generating experiment, collection of primary data, processing and analysing, publishing and sharing, preservation, and re-use of data [2]. Awareness for the challenges in long-term data management has reached a level where structural measures are put into practice, e.g. funders requiring detailed data management plans in grant applications [3]. In several scientific communities including the life sciences, further political and technological efforts to permanently install high-quality data management in the scientific process are currently supported by commitments to the FAIR Guiding Principles [4] (Findable, Accessible, Interoperable, and Reusable data). Unique and persistent resolvable identifiers (PID) [5] and descriptive metadata compatible with semantic web technologies [6] are widely applied examples of enabling tools for FAIR data sharing. Their efficient integration into routine workflows and information systems in biomedical research should be a central goal for infrastructural software development.

Several studies suggest negative economic consequences of insufficient experiment reproducibility are crucial alongside scientific and ethical aspects of RDM. Research results based on irreproducible studies might lead to a huge waste of money, staff time, and can result in delays in drug development or termination of clinical trials (e.g. [7, 8]). About 1/3 of preclinical irreproducibility results from biological reagents and reference materials [7]. Li et al. report that the description of experiment materials are key for research results, although often insufficiently reported [9]. This is especially true for antibodies [10], cell lines [11, 12], and animal models [13].

Diverse stakeholders demand researcher groups and projects to fulfil a large variety of requirements regarding RDM. Many of those requirements have to be addressed by the biomedical researchers. Amongst these stakeholders are funding agencies (good scientific practice), biomedical journals (authors guidelines), universities (data policies), core facilities (terms of use), and research group leaders (traceability).

Recommendations published by the German Research Foundation (Deutsche Forschungsgemeinschaft, DFG) in their guideline for Good Scientific Practice (GSP) [14] include the record keeping of (published) research data for at least ten years. Apart from the technological challenges with e.g. changing file formats and storage devices over such a long time, this implies requirements regarding the metadata enrichment in terms of data integrity and intelligibility. The latest revision of the GSP guideline (published in September 2019) [14] explicitly states that all materials, methods, and data that belong to scientific publications should comply with the FAIR Guiding Principles.

This also encompasses the upload of annotated research data to public repositories as required by some journals. A collection of such public repositories can e.g. be found at the Registry of Research Data Repositories [15] or the FAIRsharing registry [16].

Biomedical experiment workflows tend to be increasingly complex regarding the application of highly sophisticated and expensive measurement devices / techniques and are therefore increasingly supported by specialized service facilities [17]. Examples include technical approaches for nucleotide sequencing, advanced light microscopy and nanoscopy (ALMN) [18] or echocardiography of research animals [19]. Professional RDM to support these facilities exemplarily includes the documentation of necessary planning steps, (sometimes very large) raw data file generation and storage, complex analytical pipelines, special software products, as well as proprietary or non-proprietary raw and metadata formats.

Regarding the relatively small time-budget of an experimental scientist, the balancing between complete and sufficient effort to collect and document as much information as possible about experimental materials and workflows is challenging. Experimental documentation must be noted down in (usually paper-based) lab notebooks of the researchers performing the experiments. Moreover, if, e.g. antibodies are used by several lab members of a research group, this information should be present in every single notebook. In some research groups the switch to Electronic Laboratory Notebook (ELN) based documentation is intended to simplify and enhance some aspects of GSP and RDM [20].

These insights have been translated into a data management infrastructure developed by the subproject infrastructure (INF) within the biomedical Collaborative Research Centre (CRC) 1002 at the Göttingen Campus. This CRC investigates heart insufficiency across multiple partnering research groups. Since 2012, data management requirements of the participating researchers have been collected and transformed into an integrated web portal which supports the facilitated documentation of biomedical laboratory workflows, internal data sharing, publication of results, and transfer of data into public repositories. Here, we describe the architecture, implementation, and resulting benefits of the developed system as well as the possibility to reuse the software packages in different biomedical project settings. While the conceptual approach has been initially described before [21], a more detailed presentation of technical implementation and recent developments is summarized in this report.

### Objectives

Goals of the described system can be summarized as follows. The system shall:

- (G1) improve documentation and findability of experimental assets such as antibodies, cell lines, mouse lines;
- (G2) improve experiment documentation by collecting workflow metadata at the time of initial execution;
- (G3) enable data storage and data sharing between researchers from different domains;
- (G4) increase adherence with the FAIR Guiding Principles for all information related to published scientific articles.

All of the above shall be implemented as a system that:

- (G5) maximizes usability,
- (G6) is compliant with information security guidelines,
- (G7) is compliant with best practices regarding software sustainability, and
- (G8) is optimized for reuse in other biomedical research projects.

## Implementation

The "*menoci*" software suite (from the Latin words: **me**moria, **no**titia, s**ci**entia) described in this article consists of multiple modules, which are grouped around a central published data registry (see Figure 1). This software focuses mainly on two different aspects of RDM, i.e. the collection and integration of data, and the comprehensive documentation and workflow support of all steps from planning through to sharing and publishing. In general, four different types of modules need to be distinguished: (i) archive, (ii) registry, (iii) catalogue, and (iv) method-specific workflow support. The Research Data Archive module serves as the storage backbone for managing the different datasets (i) of the workflow supporting modules for Echocardiography, and Advanced Light Microscopy and Nanoscopy (ALMN) (iv). Published scientific articles as well as paper-based laboratory notebooks can be registered (ii). Comprehensive metadata about antibodies, mouse lines, and cell models are collected using the available catalogue modules (iii).

The *menoci* software is implemented as a set of extension modules for the Drupal web content management system and development framework [22]. Drupal is a well-established open source software project with a large and active community of users and developers. The inherent modular architecture of the framework enables plug-and-play extension of basic website functionality. Complex multi-user information and communication systems can be created without actually writing software source code or accessing the host server operating system. For a highly customized user experience with fine-grained interaction and suggestion mechanisms though, programmatic extensions may be used to achieve satisfying results quickly and without complex configuration. The *menoci* suite therefore relies on the robust and easy to use scaffolding of the Drupal web portal with user management, role and permission based access management, and content management functionality. At the same time, custom functionality is implemented to allow for simple and reproducible generation of different content types (publications, antibodies, mouse lines, etc.) as well as cross-module integration.

Some modules from the *menoci* suite implement client libraries for a specific REST service component and expose the service functionality in PHP according to Drupal framework conventions. One such library enables interaction with an ePIC PID service provider [23] to create and manipulate PIDs for use in the catalogue modules of the system. Another library enables communication with a data storage service based on CDSTAR [24] for permanent file and metadata storage used by the workflow supporting *menoci* modules.

## Functional Design

The *menoci* modules are designed for building an integrated project website and data management portal. Both affiliated researchers and public audiences can access information according to the Drupal role and permission based access model. Through sub-group configuration, permissions per project are customisable, i.e. allowing the distinction between different labs or partner organizations. Researchers can be assigned different functional roles within one or multiple groups that encompass a configurable set of permissions. For example, a person may be principal investigator for one group, including special permissions to view the group members' entered data, while at the same time only having restricted access to data from a second group. Cross-group collaboration can be documented using configurable sub-projects.

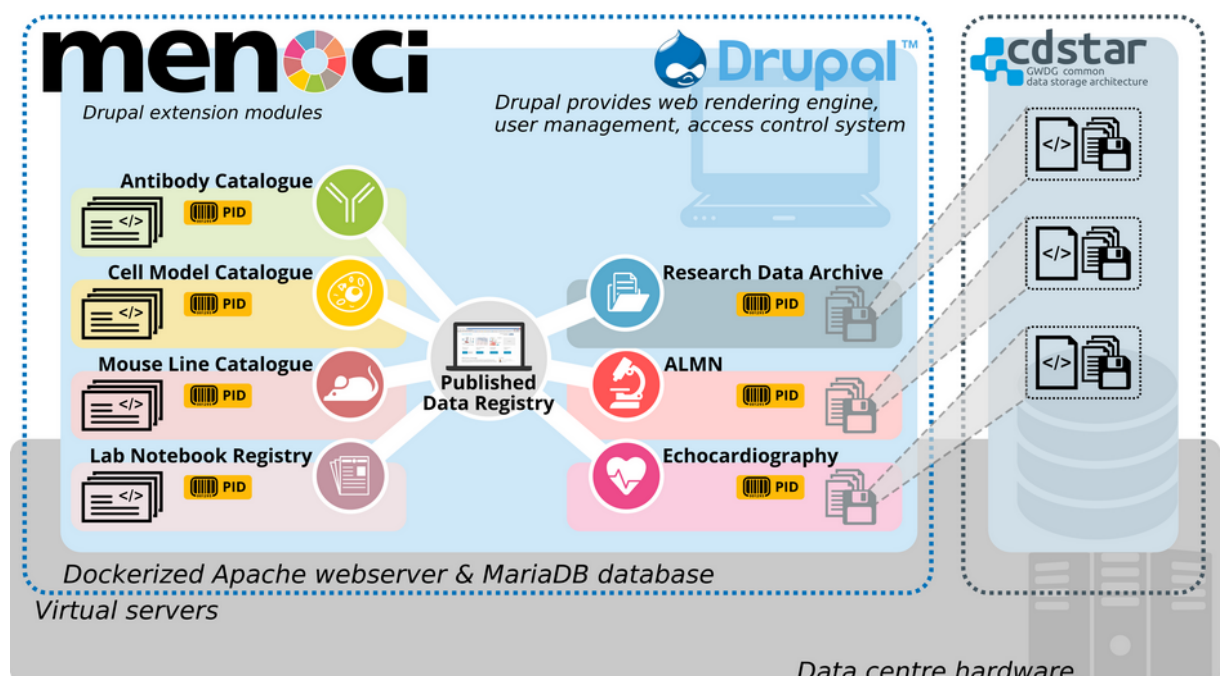

**Figure 1: System architecture schema of menoci.**
A possible deployment configuration of the menoci system is depicted wherein both the Drupal webserver and database and the independent CDSTAR storage service a run from separate virtual servers within the same data centre. Drupal installation is again virtualized using Docker container technology for ease of installation and updating. The menoci system consists of eight modules providing distinct features The Published Data Registry module is at the centre of the arrangement as it provides the integral list of scientific publications which is the entry point of linked assets and data. Antibody Catalogue, Cell Model Catalogue, Mouse Line Catalogue, and Lab Notebook Registry on the left each provide specific metadata schemas for experiment asset documentation. Research Data Archive, Advanced Light Microscopy and Nanoscopy (ALMN), and Echocardiography module on the right allow for result data storage. Data and annotated metadata are persistently stored using the CDSTAR service. All assets and data documented through menoci are assigned persistent resolvable identifiers from the Handle system. Menoci icons have been created using images from Flaticon and Freepik that are licensed CC-BY 3.0.

Each registered user of the system may log in using the specified credentials. By default, Drupal supports username/password based authentication but may be altered using a wide range of publicly available extension modules. Coupling with organisational identity management or federated identity management systems is possible using LDAP, SAML, or OpenID extension modules. The *menoci* base module automatically creates a user profile containing family and given name, and the ORCID ID. Profiles can be arbitrarily extended and accessibility restricted according to privacy policy. Name and ORCID ID are required for visual and semantic markup of user information by *menoci* modules though.

To facilitate data sharing and reuse, collected information can be exported in structured formats and, where applicable, is presented with semantic annotation. In the current version of the *menoci* system, basic semantic data about scientific articles compliant with the schema.org/ScholarlyArticle type definition [25] is integrated into the HTML of human readable web pages as JSON-LD objects. By changing the HTTP request header "Accept" to the value "application/json", this object can be directly retrieved, creating a simple web service API providing machine readability.

**Archive module**

Data storage (i.e. for uploaded files) is realised using the CDSTAR software developed by the campus scientific data centre Gesellschaft für Wissenschaftliche Datenverarbeitung Göttingen (GWDG). CDSTAR is an open source package oriented storage and archiving service for binary files and metadata [24]. A REST web service API is exposed that provides an abstraction from underlying storage media. Data sent to the service are processed and stored in defined logical domains along with individual or role-based permissions and metadata. The service enforces ACID transactions [26] and supports configuration to switch between hot and cold storage, e.g. to move least-recently used large files from fast and highly available to less expensive storage media fit for long-term archiving. While similar to popular *object* storage services like Ceph or Amazon S3, the *package* oriented CDSTAR allows clients to store multiple related files in one place. Such packages are accessible by ID and include specific metadata for each file.

By using CDSTAR as a dedicated back-end storage service, researchers using the *menoci* system are presented with an easy to use tool to not only keep their data accessible and findable for their own needs but readily stored for long-term preservation and data sharing. Data stored in CDSTAR is only accessed through the user interface provided by *menoci* modules using a 4-level role-based access model: data and information can generally be set to be publicly accessible, accessible by members of the project (i.e. registered users of the Drupal instance), accessible by members of the same research group (as defined by roles assigned through Drupal user management), or privately accessible for the owner/creator of the data item.

Globally-unique persistent identifiers from the ePIC consortium are assigned to data registered or published through the *menoci* platform. The registered PIDs are compliant with the *Handle* system [27] which is also the underlying technical and organisational infrastructure of the widely applied Digital Object Identifiers (DOI). A back-end module for communication with the ePIC API has been implemented that lets administrators configure one or multiple service endpoints and in turn assign these to specific *menoci* modules to be used for PID registration. PIDs are currently registered either automatically when inserting information about a new item into one of the catalogue modules, when creating printable barcodes and unique labels as part of the ALMN workflow module, or manually triggered by a user actively deciding to register a PID for a set of uploaded data. Each PID is registered with a target URL linking back to a publicly accessible page of the Drupal instance. Such public landing pages are designed to display minimal metadata about the registered logical object associated with the PID even if the actual data is set to have restricted access by the data owner. The Lab Notebook Registry, Mouse Line Catalogue, and Antibody Catalogue modules within the *menoci* suite currently still require manual pre-registration of PID sets by the project team. Refactoring of these modules is required to switch to an automated on-demand PID registration.

**Registry modules**

Published Data Registry

The core of the *menoci* suite is the "published data registry", a feature-enhanced list of scientific publications created by members of the project. Article metadata can be registered by manual data entry, or more conveniently by querying data from the publicly accessible EuropePMC [28] and DataCite [29] API either via a PubMed ID or DOI assigned to a specific article. Items are cross-linked with research groups and subprojects. The (sub-) type of each article is assigned according to the respective PubMed vocabulary [30]. Arbitrary external resources that can be identified by URL may be configured and added to an item. The list of scientific articles can be searched, filtered, and exported. Available export formats are RIS, JSON, and spreadsheet (XLSX). Basic descriptive statistics like the number of publications per year or ratio of open access versus non-open access publications can be displayed and visualised using the Plotly library [31].

The primary concept for cross-module relationship in between available data within a *menoci*-based system is linking information about assets to a related scientific article. For each scholarly article, related items from other *menoci* modules can be marked as associated and will be displayed as such. It is possible to link registered laboratory notebooks, antibodies, mouse and cell lines, microscopy and echocardiography experiments, and uploaded data files to an article.

### Lab Notebook Registry

The Lab Notebook Registry module provides users with the option to create a digital representation of their original paper notebooks used for primary experiment documentation at the bench. A pre-registered PID can be coupled with barcode stickers to create a durable connection between the physical item and its digital counterpart. To ensure unique application of barcode stickers, an auxiliary one-time transaction code (TAN) must be entered when registering a lab book. For registered notebooks, files may be uploaded, e.g. to store a scanned page or the scan of an entire lab book as a digital copy of the original. To support the physical storage and archival process, a storage location may be specified as part of the descriptive metadata collected during registration.

## Catalogue modules

### Mouse Line Catalogue

The Mouse Line Catalogue allows researchers to store and share mouse related research data and represents an exemplary approach to the standardized documentation of model organisms. It offers the possibility to document single mouse lines and associated individual mice. The collected information comprises details about the mouse line itself such as breeding type, strain name, the originating laboratory and Mouse Phenome Database identifier [32] as well as information on genetic mutations such as mutation type, name and abbreviation of the mutated gene according to the gene nomenclature of the NCBI Gene Database [33], and provenance data. In addition, individual mice can be added to a created mouse line entry with information about name, sex and birth dates.

The main characteristic incorporated in the mouse line catalogue is the dynamic name generation of a mouse line while information is entered by the users in respective input masks. The name generation is primarily based on the rules and guidelines established by the International Committee on Standardized Genetic Nomenclature for Mice [34].

### Cell Model Catalogue

The main function of the cell model catalogue module is to aid researchers in the documentation of induced pluripotent stem cell (iPSC) lines by providing interactive well-structured forms for entering the respective metadata for each cell line. Supported are patient-derived cell lines as well as genetically modified cell lines. Besides general information about the cell lines, the documentation is focused on information about the cell culture, donor, diagnosis, ethics and biological verification. An optional but recommended step of the documentation process is to generate a standardized name for the cell line via the naming tool API of the Human Pluripotent Stem Cell Registry (hPSCreg) [35]. The gathered data are searchable and are represented in a structured way for both internal and external users. Data may also be imported or exported via a predefined spreadsheet format.

### Antibody Catalogue

The Antibody Catalogue offers researchers the possibility to document a broad harmonized set of metadata about antibodies and link this information directly to publications. Entering and exporting data takes place in structured forms, which comprise information about antibodies as well as applications (e.g. Immunofluorescence, western blot). Additionally, researchers can enter anonymous assessment of how well a specific antibody has worked in one of these applications and add images of the process and results. Optionally, selected data may be exported as well as imported via a predefined spreadsheet. The antibody catalogue module is divided into two main parts: primary antibodies and secondary antibodies. For each antibody, information is stored about the manufacturer from whom the antibody originated. Furthermore, it is possible to link items with external sources like Antibody Registry [36] and Antibodypedia [37].Items stored in the catalogue can be found using the search function if the creator of the antibody dataset released it with the necessary access permissions.

Apart from the ability to connect mouse lines, cell lines, and antibodies to scientific articles listed in the Published Data Registry, cross-linking of such assets with specific experiments within the workflow supporting modules is planned. In the current feature set, antibodies used for staining in microscopy experiments within the *menoci* ALMN module are referenced from the antibody catalogue.

**Workflow support modules**

Advanced Light Microscopy Nanoscopy (ALMN)

The ALMN module is designed as workflow support for microscopy experiments and to enhance reproducibility and subsequent reuse of microscopy derived datasets. The module supports experiments from the conceptual planning phase until storing of the generated datasets and integrates image data with associated metadata. Along the organisational workflow of the CRC 1002 microscopy service project (see Figure 2), researchers have to input the data in multiple steps during microscopy experiments. The first input step asks for general information about the researcher and the planned experiments, e.g. the research question and the planned procedures. Second is a "consultation" step that may include a meeting with service project ALMN experts depending on the experience of the researcher. In the consultation step, project metadata like information about the applied stainings (e.g. immunofluorescence stainings) and samples is gathered. References to the Antibody Catalogue module enforce standardized descriptive antibody information. The module collects information necessary to create physical labels and offers these attributes like sample ID, species and staining abbreviation, and PID as a downloadable list. In our primary setup at the CRC 1002, a commercial label-printing system is processing this data. The printing system produces adhesive labels with barcodes containing PIDs to link the physical coverslips to the collected (meta-) data.

Following the actual data acquisition at the microscope, the researcher may upload the generated (image and meta-) data as a Zip-compressed file via the web interface. Zip files are used as a transfer format to preserve the folder structure of the generated data, which is necessary for our automated metadata extraction. Finally, experimental metadata, microscope device metadata, and the image data are stored together and are accessible via the *menoci* platform. The datasets can be shared with other research groups or made publicly available according to the *menoci* access rule system described above.

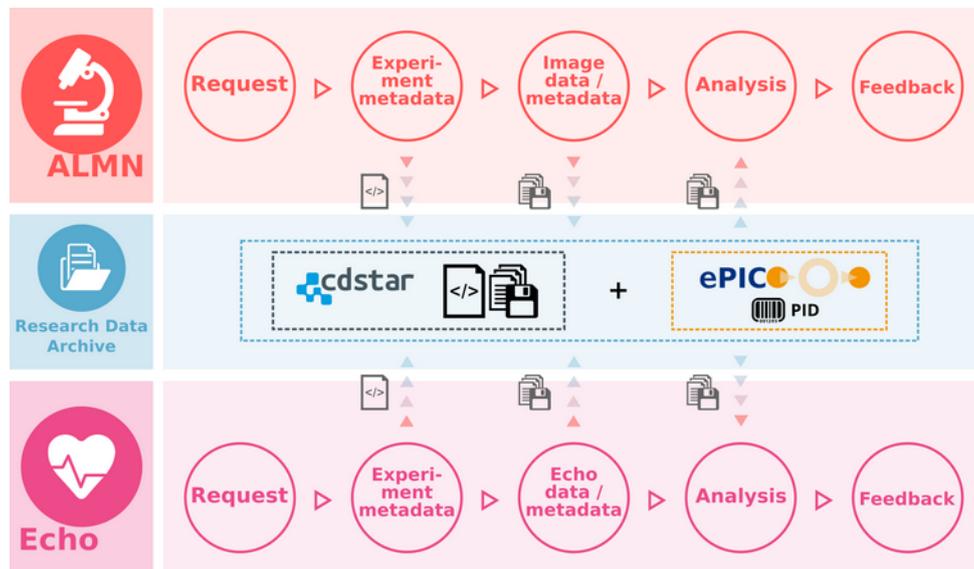

**Figure 2: Schematic representation of the menoci ALMN and Echocardiography module workflows.**
The menoci Advanced Light Microscopy and Nanoscopy and Echocardiography modules guide researchers through the workflows of CRC1002 service facilities for the name-giving imaging methods. The software tracks status of different service requests by researchers, presenting customised data entry forms for all workflow steps and streamlining communication with facility personnel. Microscopy or Echocardiography project metadata and resulting files are stored using CDSTAR via the menoci Research Data Archive module, and assigned Handle persistent identifiers (PIDs) using the ePIC web service. Researchers and facility staff can retrieve datasets for further analysis and finally may use ALMN and Echo module functions to communicate feedback. Menoci icons have been created using images from Flaticon and Freepik that are licensed CC-BY 3.0.

Echocardiography

The Echocardiography module supports researchers in the process of planning, executing, and evaluating echocardiographic image acquisition. Similar to the ALMN module, researchers can request echocardiographic imaging performed by a CRC 1002 service project (see Figure 2). Initially, researchers need to provide specific descriptive information about mice, surgery type and the echocardiographic imaging experiment design (e.g. time line). The responsible echocardiography service project specialists review, accept, and execute the request or may in turn demand further information. Along the workflow progress, relevant data and metadata are collected. While some information has to be entered manually by the users, most of the information is automatically extracted from the data generated by the imaging devices. Upon uploading, the imaging datasets are linked to the respective experiment records in the Echocardiography module. An evaluator may be assigned to a study, who can process the result data as soon as an experiment is marked as finished.

## Software Development

The software development process behind the *menoci* modules follows the agile development paradigm [38]. Beginning with the project start in 2012, *menoci* developers have been continually working with designated users from biomedical research groups and labs to adjust development based on user expectations and experience. Through this interaction, topics that would provide major positive impact on research documentation were pushed into development early. The *menoci* source code is hosted and maintained in an open GitLab instance provided by the local academic IT-service provider GWDG. Each Drupal module is tracked as a dedicated git repository within a project-centric group section. The GitLab issue tracking tools are used to structure tasks, sprints, and releases. Sprints are organized regularly yet not in intervals as short as proposed by popular guidelines due to the sparse resources of the team that consists of two part-time computer scientists and multiple student research assistants.

# Results and Discussion

The features described above make a fully configured instance of the *menoci* system a valuable asset in biomedical research data management. Considering the steps of the RDM life cycle, major contributions to the collection, publication, analysis, sharing, preservation, and reuse of scientific data can be achieved. In the following, we discuss these contributions from a researcher's perspective (in line with objectives G1-G5), from an infrastructure provider perspective (objectives G6-G8), and from a technological/developer point of view. We are briefly describing already planned features and potential future developments. Finally, we present an overview of available alternative systems and compare scope and main features with the *menoci* suite.

## Researcher Perspective

### (G1) Experimental asset documentation

Central aim of the *menoci* web portal is the support of standardized high quality documentation of experiments and derived data along the scientific process as well as an easy-to-use researcher interface to cross-link and publish relevant datasets unambiguously with the manuscripts in scientific journals.

*Menoci*'s Published Data Registry enables rapid integration of bibliographic information about scientific publications via the PubMed and DataCite API by entering PubMed ID or DOI. Subsequently, experimental assets related to the publication can be associated from the available catalogue or registry modules. Prerequisite is the registration of assets in our system, which in turn necessitates transfer of material management information (e.g. spreadsheet- or database-based) into *menoci*. For the projects supported by our group, lists of laboratory assets have been collected from the participating research groups, harmonized to the common metadata model, and imported into the system. Currently, batch import via Excel-spreadsheets is available for the Antibody and Cell Model Catalogue to facilitate the integration of large collections. Harmonizing and centralisation of asset information is essential to reduce issues caused by versioning and integrity of the collection management. By registering globally unique persistent and resolvable identifiers via the Handle system, assets can be unambiguously referenced across articles and projects, strengthening reproducibility of described experiments.

### (G2) Experiment documentation

Enforced, standardised documentation and reference of experiment assets are enablers for overall improved documentation of biomedical experiments. Still, whether this results in actual benefit depends on the individual researcher's style of documentation and if the unique identifiers for assets are actually applied. To further streamline documentation of widely standardized experiments, *menoci* modules offer workflow support for microscopy and echocardiography studies. The workflow support design allows data to be collected at the moment of generation. Utilized assets can directly be selected from the available catalogues, facilitating unambiguous references. Finally, result data may be stored in form of microscopy images and metadata, or echocardiography measurements and parameters. *Menoci* enables for complete documentation of experiment workflows from the initial research question via experiment design to resulting datasets including participating user information and audit trail functionality.

Such workflow documentation tools need to be tailored to individual needs to generate maximum benefit in routine experiment settings. This is why the ALMN and Echocardiography modules are customised towards local service project workflows within CRC 1002. Application of these modules in new environments might necessitate adaptations in web portal design (e.g. user guidance, database structure). However, the modular *menoci* design facilitates this process. We currently support the pilot adoption of the ALMN module in a second research project.

The ALMN module automatically extracts rich metadata from uploaded files, enhances searchability and reproducibility by harmonizing semantical annotation. This requires laborious case-by-case analysis and processing of the image and metadata formats produced by different devices. The ALMN module microscopy data storing functionality is per se agnostic and therefore independent of specific file formats. But in order to maximally benefit from automatic extraction and representation features, the images and microscope metadata should be uploaded as tiff and xml (or any other text-based file) format, respectively. Due to our observation that image analysis pipelines and applied tools have a high degree of heterogeneity, we decided to exclude analytic tool integration or development from *menoci* design. However, to gain complete experiment documentation there is an obvious need to cross-link corresponding bioinformatics workflow documentation with the ALMN datasets. Concepts are currently designed and evaluated.

Support for paper-based experiment documentation is offered by the Lab Notebook Registry module. Unambiguous reference of laboratory notebooks allows novel connection of a scientific article with the offline experiment documentation. Here, the *menoci* suite is the link between the physical paper-based lab notebook and the digital (meta-) data. Managing and distributing barcode stickers and TAN lists to researchers creates organisational overhead that cannot be solved entirely in software, though.

**(G3) Data storage and sharing**

Storing research data in a *menoci*-based web platform with a fully configured CDSTAR storage backend and PID registry facilitates improved data sharing within projects and research groups. This approach implements important aspects of the FAIR Guiding Principles beyond the requirements of Good Scientific Practice. Technological and house-keeping efforts of managing raw files containing research data can be reduced when using the Research Data Archive as primary storage location. Preservation of data as well as publication and sharing are supported by the *menoci* modules through the common access control system described above.

Furthermore, collecting data about experiment assets in a harmonized rich metadata schema facilitates sharing of the information with public repositories. Antibody data can e.g. be exported and transferred into the public knowledge base Antibodypedia, cell model data can be automatically registered in hPSCreg by communicating with the API. Sharing asset data in public repositories may be of help to fulfil journal requirements along scientific publishing.

**(G4) FAIR guiding principles**

Using a *menoci*-based system to manage assets and data of a biomedical project can help to improve FAIRness. Comparing the implemented features of the systems against the first published version of the FAIR Metrics [39, 40], compliance with a subset of the FAIR criteria can be achieved: Utilization of Handle system PIDs for assets and datasets, display and download of machine-readable metadata, and use of the open HTTP(S) protocol are important measures to achieve findability and accessibility. Data access control is consistently enforced through the Drupal framework. Some criteria focus on documentation and implementation on a per-instance basis rather than on functionality of the software: A contingency plan for resource availability should be available on project basis as well as formal documents describing the implemented formats, protocols, and measures. Templates for these documentation artefacts might be provided by the developers of the software system but are not yet available in this case. Interoperability and reuse of data are partly supported where semantic JSON-LD markup of collected information is accessible or data is replicated into public data repositories like hPSCreg or Antibodypedia.

Expansion of the JSON-LD API for semantic data retrieval is under active development. Nested information about related items from the *menoci* catalogues are integrated into the schema.org/ScholarlyArticle [25] JSON object as a next step. This will enable retrieval of relational maps between semantic objects on a granularity level that no other scientific information source on the web can currently match.

In summary, compliance with "findable" and "accessible" criteria is well progressed for assets and datasets that are stored using *menoci* modules and are declared publicly visible by the creators. The "interoperable" and "reusable" criteria may be assisted but are not sufficiently supported by documentation templates or software features yet. As the FAIR Metrics definition process continues, future work on the *menoci* suite should follow the route to further support out-of-the-box compliance with the entire set of criteria.

### (G5) Usability

The Drupal content management system "Theme" engine is used to markup all display and interactive dialogue as HTML web pages. Presentation as web pages enables platform-independent usage. To maximize accessibility of the *menoci* services for users both at the desk and in the laboratory, the *menoci* modules require a Drupal theme based on Bootstrap [41]. Bootstrap provides full content responsiveness.

Data entry is achieved through interactive web forms created by the Drupal form engine. Inserted data can be validated given basic or advanced custom validation functions. Forms required for data entry within the *menoci* modules provide users with pre-defined options for all information that may be collected in structured format. Tooltips are displayed to the user when editing free text items.

Usability of the modules is in summary designed by the development team in constant feedback with actual users. Still, this is a best-effort basis, which could be improved in the future by consulting with user interface and user experience experts.

## Infrastructure Perspective

### (G6) Information security

Securing confidentiality, integrity, and availability of data managed using *menoci* modules is largely a task that must be specifically regarded per system/instance deployed. Routine integrity checks of the data, backup, and system availability must be implemented individually. Security measures to prevent unauthorized access and manipulation of collected data are implemented by the Drupal framework itself. By relying entirely on the form and access control engines of Drupal, *menoci* modules delegate most of these measures. Known vulnerabilities for Drupal are publicly tracked [42] in compliance with the Common Vulnerabilities and Exposures system [43]. Vulnerabilities are regularly addressed through software updates and patches. Drupal has been assessed positively regarding security among popular CMS software alternatives [44]. Regular installation of updates to the core components must be performed by administrators.

### (G7) Software sustainability

Software sustainability refers to the complex task of maintaining correctly functioning code and application over the course of an entire software life cycle. A lack of evaluation methods has been determined and software produced and operated in a research context is regarded as an especially challenging subject [45]. Simple measures to improve sustainability of research software projects have been proposed, i.e. to make source code openly available, to publish descriptive metadata, to adopt a code license, and to define contribution and communication processes [46]. The primary measure to support sustainability of the *menoci* software project is publication of all source code under the open GNU General Public License. Following software citation suggestions [47] we provide metadata in JSON-LD format, keep an updated list of authors and contributors to the project, and describe the process of contributing code. The question of how to automatically provide persistent identifiers for source code and code versions is not yet solved [48]. The initial distribution version of the *menoci* source code is replicated into the Göttingen campus repository (doi:10.25625/NC9TF6), a list of code versions, metadata, and assigned DOI is kept at the project website https://menoci.io. This current, manual process does not scale, is prone to error, and likely to fade quickly due to regular changes of responsibilities and context in research institutions. The automation of DOI or Handle minting triggered by version release holds strong potential for improvement.

**(G8) Reuse in other projects**

Several design choices of the *menoci* project may facilitate reuse of the system in different biomedical research projects. Extending a well-established web CMS is in line with proposed methodology for software development in lab and research group context [49]. Implementation of the system as a set of loosely coupled extension modules for Drupal allows for simple deployment, composition, and integration. The distribution package and documentation at the *menoci* website are specifically tailored to enable minimal-effort deployment of a fully functional instance for exploration purposes. Combining a subset of the *menoci* modules with external Drupal extension modules enables integration with existing platforms, for example by using single sign-on mechanisms and exposing only the frontend of the desired *menoci* module while hiding all further Drupal functionality from the end user.

**Technological Perspective**

From a technological software engineering perspective, the *menoci* project is a work in progress rather than a finished product. The modular architecture strengthens maintainability of the source code, i.e. allowing for incremental refactoring of individual modules. A major challenge on the horizon is the planned end-of-life of the underlying Drupal version 7 in November 2021 [50]. While commercial extended lifetime support or even a fork of the Drupal 7 code base may be viable options for active instances of *menoci*-based platforms, migration to the latest Drupal major version is recommended [51]. However, the migration from Drupal version 7 to 8 is labour-intensive as the framework has been completely refactored and a step-by-step migration plan for the *menoci* modules is required given the currently limited software development resources. Incremental migration starting with core modules is a viable approach but not yet decided upon or scheduled as an official strategy.

When applying a software development framework like Drupal, a major design decision is the level of strictness of adhering to the framework's core design rules as well as boundaries. Basing *menoci* entirely on Drupal's core design concept as a module centred software maximizes compatibility with third-party modules and enables essential general functionalities like user and administrative management. On the other hand, relying less on Drupal's core modules allows for more flexibility in design and reduces development time for specialised modules. Considering for example the Published Data Registry, which is essentially a list of scientific articles: Creating lists of things is a standard use case of the Drupal content management system. It is possible to create a custom content type with all specific attributes entirely through configuration of the CMS as admin user. At the same time, this functionality may be built entirely in source code while just using Drupal as a canvas with routing capabilities, form rendering, user management, and access control. The trade-off is between being able to extend the available attributes of the content type without having to write code and being able to custom-tailor forms, validations, and interactive user experience as well as integration with external services exactly towards requirements. The *menoci* project currently follows the second path. Especially in the early project stage, this choice enabled quick development progress and prototyping of functionality. Among popular CMS, Drupal has been judged best suited for professional software development [52]. Integration of self-configured content types with *menoci* modules is a likely future feature to enable even more flexibility for reuse of the project. Facing the refactoring required for migration to a newer Drupal major version, we will evaluate the possibility of gradually moving towards Drupal API compliance.

**Future Work**

Although *menoci* offers basic and even specialised modules for RDM, possibilities for the extension and improvement are manifold. Apart from expansion to further increase functionality and usability for researchers, quality assurance and integration with third-party tools and systems are dimensions of future work. One key quality aspect is the technological sustainability of the *menoci* source code. Refactoring of the code base and expansion of the active developer community need to be considered. While the *menoci* system is currently rolled out in multiple projects, the development effort is still fully centralized at one medical informatics research group. Launch of a project website and dissemination among the scientific community are first steps already taken towards spreading development among multiple stakeholders. Survival of the project at this critical stage of the software lifecycle depends on available resources and support at the core research group.

Regarding the integration with third-party software and systems, any combination of systems and tools is conceivable that may improve data management workflows for researchers. Integration with popular third-party systems in use at many different sites might increase the value of *menoci* modules for reuse in different projects. For example, interfacing with the research data repository software Dataverse [53] would not only facilitate an integration of systems deployed at the Göttingen campus but immediately allow other sites that also rely on an instance of Dataverse to fit *menoci*-based platforms into an existing data management ecosystem. At the same time, site-specific integrations may significantly improve workflows for local researchers within a given project. Current effort to interface the Cell Model Catalogue module with the UMG Biobank information systems is an example where benefit of the development is exclusive for the local setting. Developing the *menoci* system to seamlessly integrate with site-specific research data infrastructure [54] as well as emerging national [55] and global ecosystems [56] is a major strategic challenge.

Since its launch in 2012, the development of the *menoci* platform was mainly driven by the extension of features and optimisation of user experience. Improving performance and software quality should become a focal point given continued usage of the system in multiple research projects and groups.

## Alternative tools

Looking at typical requirements for a research data management system in a biomedical research project, the list of features extends multiple categories of applications. Laboratory information management systems (LIMS) manage laboratory orders and processes, as well as results of ordered laboratory analyses. Data repository systems store and manage datasets, enabling publication and search using standardised metadata. Content management systems (CMS) focus on content distribution project- and worldwide.

The *menoci* approach merges these aforementioned features typically covered by content management systems (CMS, laboratory) information management (LIMS), and data repository systems into a single modular system.

Many specialised tools exist, which only cover one of these categories. Especially all-purpose CMS systems, which are not bound to the scientific world, are available in a wide selection. The dominant WordPress besides Joomla and Drupal are examples for popular choices when creating dynamic websites and web portals. CMS provide generic content management functionality and may serve to fulfil requirements regarding data publication and laboratory asset management. LIMS are generally specialized in laboratory process and sample management, covering requirements regarding experiment documentation and workflow support. LIMS are dominated by commercial software and target drug manufacturing laboratories as the main audience. Senaite [57] and OpenBIS [58] are examples of open source LIMS implementations with OpenBIS providing additional electronic lab notebook capabilities [59]. With a big push to unify and increase availability of data in biomedical research, the supply of research data repository software is increasing and especially rich in open source solutions. Core functionalities are upload of datasets, annotation with metadata, and subsequent publication of both following linked data principles. Some popular examples are Dataverse, DSpace [60, 61], Fedora [62, 63], E!DAL [64], and Invenio [65].

Open source alternatives that cover the whole range of functionality described above are scarce and have been even less established when development and use of *menoci* started in 2012. Solutions that might be configured and extended towards a similar functionality as a *menoci*-based platform are SEEK and VIVO. SEEK is an academic software project that can be deployed locally and strongly emphasises research project data management and published article-related data sharing [66]. The ISA model [67] implemented in SEEK supports semantic experiment description and documentation. Functionality for asset catalogue management is limited though. A centralized SEEK installation for open data exchange is available as FAIRDOMHub [68]. VIVO [69] is an ontology-based scholarly information management system originally focussed on creating a web-based representation of research organisations with department hierarchy and researcher profile pages. Since all user interfaces and collectable data types are defined by a configurable ontology, VIVO may be extended to cover functional aspects of the catalogue and registry components of *menoci*.

In summary, to our knowledge no alternative software can currently cover the unique set of features implemented by the *menoci* modules. The composition of features implemented by the *menoci* suite combined with the extensible Drupal framework creates a unique toolbox. We argue that *menoci* offers a valuable extension to the ecosystem of available open source software for research projects.

## Conclusions

Biomedical research projects deal with data management requirements from multiple sources like funding agencies' guidelines, publisher policies, discipline best practices, and their own users' needs. We describe these functional and quality requirements based on many years of experience implementing data management for the CRC 1002 and CRC 1190 projects and propose the *menoci* system as a solution. Source code and distribution packages of the *menoci* modules are published under open source GNU General Public License. The *menoci* modules are based on the Drupal content management system enabling lightweight deployment and setup, and create the possibility to combine research data management with a project home page or collaboration platform.

Management of research data and digital research artefacts is currently in the middle of a transformation process from manual housekeeping of local files to project or organisational managed services which in turn will contribute to overarching data sharing infrastructures [55, 56]. To enable and support this structural transformation process, a vital ecosystem of open source software tools is needed. The *menoci* suite is designed to be a contribution to this ecosystem but depends on continuation of development resources and a wider community of users and contributors.

## Availability and requirements

All menoci modules described in this article are available for inspection, use, and improvement as defined per GPL. The official distribution through the project home page consists of the core libraries and the following extension modules: Published Data Registry, Research Data Archive, Antibody Catalogue, Mouse Line Catalogue. Source code of further menoci modules is publicly available at the project GitLab repositories. Modules that are not yet part of the official distribution package require code refactoring to remove site- or project-specific branding or functionality.

For installation and quick start instructions, please refer to the project home page.

**Project name**: menoci

**Project home page**: https://menoci.io - Handle to refer to the project source code repository: https://hdl.handle.net/21.11124/menoci-0001 - DOI referring to menoci project 1.0 release version: 10.25625/NC9TF6

**Operating system(s)**: Linux-based OS are recommended for webserver and database. Server-side functionality can be fully virtualized using Docker images provided by the menoci project.

**Programming language**: PHP (currently supported version: 7.2)

**Other requirements**: An instance of the Drupal 7 CMS is required to host the menoci extension modules. To use full functionality of the Research Data Archive module, an instance of CDSTAR storage service as well as an account with a Handle PID service provider are required. Compatibility with GWDG and surfSARA PID services from the ePIC consortium is currently implemented.

**License**: GNU General Public License 3.0

# List of abbreviations

ACID Atomic Consistent Isolated Durable
ALMN Advanced Light Microscopy and Nanoscopy
API Application Programming Interface
CDSTAR Common Data Storage Architecture
CMS Content Management System
CRC Collaborative Research Center
DFG Deutsche Forschungsgemeinschaft (German Research Association)
DOI Digital Object Identifier
ELN Electronic Laboratory Notebook
ePIC Persistent Identifier Consortium for eResearch
FAIR Findable Accessible Interoperable Reuseable
GNU GNU is Not Unix
GPL General Public License
GSP Good Scientific Practise
GWDG Gesellschaft für Wissenschaftliche Datenverarbeitung Göttingen
hPSCreg Human Pluripotent Stem Cell Registry
HTML Hypertext Markup Language
HTTP(S) Hypertext Transfer Protocol (Secure)
ID identifier
iPSC induced pluripotent stem cell
JSON-LD JavaScript Object Notation Linked Data
LDAP Lightweight Directory Access Protocol
LIMS Laboratory Information Management System
NCBI National Center for Biotechnology Information
NGS Next Generation Sequencing
ORCID Open Researcher and Contributor ID
PID Persistent identifier
PHP PHP: Hypertext Preprocessor
RDM Research data management
REST Representational State Transfer
RIS Research Information Systems file format
SAML Security Assertion Markup Language
TAN transaction number
UMG University Medical Center Göttingen
URL Uniform Resource Locator
XSLX Microsoft Excel XML spreadsheet format

# Declarations

## Funding

This project was funded within the INF project of the Collaborative Research Center (CRC) 1002 and the Z project of CRC 1190, both funded by the German Research Foundation (DFG).

## Authors' contributions

Contributions to this article and software development according to CRediT Contributor Roles Taxonomy: MS: Writing – original draft, Investigation, Conceptualization, Software, Visualization. CL: Writing – original draft, Investigation, Software. CRB: Methodology, Writing – review & editing. TB: Methodology, Writing – review & editing. CK; Investigation, Writing – review & editing. LKF: Investigation, Writing – original draft, software. BÖH: Software, investigation. CH: Investigation, Writing – original draft. GA: Data curation. LKK: Data curation. SR: Writing – original draft. LW: Software, Investigation. BM: Investigation, Conceptualization, Methodology, Software, Visualization. MH: Methodology, Software. PW: Conceptualization, Supervision. HK: Writing – original draft, Writing – review & editing, Conceptualization, Investigation, Data curation. US:

Conceptualization, Writing – review & editing, Supervision. SYN: Writing – original draft, Writing – review & editing, Conceptualization, Supervision.

## Acknowledgements

We thank all (former) colleagues and collaborators who have contributed to the conceptualisation and development of the menoci system.

The following third-party libraries are used in menoci source code: PHPExcel, Wikidata PHP library by Aleksandr Statciuk, FPDI by Setasign, TCPDF by Nicola Asuni, Select2 by Kevin Brown and Igor Vaynberg, Ploty by Ploty Inc.